\newcommand{\extended}[1]{}

\def \VersionWithComments{}

\PassOptionsToPackage{dvipsnames}{xcolor}

\documentclass[sigconf,nonacm]{aamas} 

\usepackage{balance} 

\usepackage{soul}
\usepackage{url}
\usepackage[utf8]{inputenc}
\usepackage{graphicx}
\usepackage{amsthm}
\usepackage{booktabs}
\usepackage{algorithm}
\usepackage{algorithmic}
\usepackage{latexsym}
\usepackage{enumitem}
\usepackage{color}

\usepackage{amsfonts}
\usepackage{stmaryrd}
\usepackage[dvipsnames]{xcolor}
\usepackage{xfrac,xspace}
\usepackage{marginnote}
\usepackage[normalem]{ulem}
\usepackage[super]{nth}
\usepackage[nameinlink]{cleveref}
\usepackage{multirow}

\setlist[itemize]{leftmargin=5mm}

\usepackage{listings}
\usepackage{tikz}
\usetikzlibrary{automata, arrows.meta, positioning,shapes.multipart}
\graphicspath{{figures/},{imgs/}}  
\usepackage[T1]{fontenc}

\ifdefined \VersionWithComments
  \newcommand{\wj}[1]{\textcolor{purple}{\marginX{}[\textbf{WJ}: #1]}}
  \newcommand{\ja}[1]{\textcolor{WildStrawberry}{\marginX{}[\textbf{Jaime}: #1 ]}}
  \newcommand{\lp}[1]{\textcolor{green!70!black}{\marginX{}[\textbf{Laure}: #1 ]}}
  \newcommand{\wop}[1]{\textcolor{PineGreen}{\marginX{}[\textbf{Wojciech P.}:
  #1 ]}}
	\newcommand{\teo}[1]{\textcolor{RoyalBlue}{\marginX{}[\textbf{Teofil}: #1 ]}}

\else
  \newcommand{\wj}[1]{}
  \newcommand{\ja}[1]{}
  \newcommand{\lp}[1]{}
  \newcommand{\wop}[1]{}
  \newcommand{\teo}[1]{}
\fi

\colorlet{colorAttribute}{PineGreen!75!green!85!black}
\colorlet{colorCondition}{RedOrange}
\colorlet{colorTransition}{Blue}

\newcommand{\eg}{e.g.\xspace}
\newcommand{\ie}{i.e.\xspace}

\Crefname{section}{Sec.}{Secs.}
\crefname{section}{section}{sections}
\Crefname{subsection}{Sec.}{Secs.}
\crefname{subsection}{section}{sections}
\Crefname{equation}{Eq.}{Eqs.}
\crefname{equation}{equation}{equations}
\Crefname{definition}{Def.}{Defs.}
\crefname{definition}{definition}{definitions}
\Crefname{algorithm}{Alg.}{Algs.}
\crefname{algorithm}{algorithm}{algorithms}
\crefname{table}{table}{tables}
\Crefname{figure}{Fig.}{Figs.}
\crefname{figure}{figure}{figures}

\newcommand{\marginX}{\marginnote{\huge{\quad\textbf{!}\quad}}}

\makeatletter
\renewcommand{\paragraph}{\@startsection{paragraph}{4}{0pt}%
  {.8ex plus 0.2ex minus 0.2ex}%
  {-0.5em}%
  {\bfseries}}
\makeatother

\makeatletter
\newcommand\dontbreakpar{\par\nobreak\@afterheading}
\makeatother

\def\event/{action}
\def\Event/{Action}

\newcommand{\A}{\ensuremath{A}}

\newcommand{\PTAMAS}{PCAMAS\xspace}

\newcommand{\MASsymb}{\mathcal{S}}

\newcommand{\model}{\mathit{M}}

\newcommand{\State}{\ensuremath{s}}

\newcommand{\prop}[1]{\ensuremath{\mathsf{{#1}}}}

\newcommand{\coop}[1]{\langle\!\langle{#1}\rangle\!\rangle}

\newcommand{\Probpath}{\mathbb{P}}

\newcommand{\Sometm}{\mathtt{F}\,}
\newcommand{\Always}{\mathtt{G}\,}
\newcommand{\Until}{\,\mathtt{U}\,}
\newcommand{\Release}{\,\mathtt{R}\,}

\newcommand{\seq}{\rho}

\newcommand{\strat}{\sigma}
\newcommand{\stratstyle}[1]{\ensuremath{\mathrm{#1}}}

\newcommand{\outcome}{\mathit{out}}

\newcommand{\irp}{\stratstyle{irp}\xspace}
\newcommand{\irP}{\stratstyle{irP}\xspace}

\newcommand{\lan}[1]{\ensuremath{\mathbf{#1}}\xspace}
\newcommand{\LTL}[1][]{\lan{LTL_{\stratstyle{#1}}}}
\newcommand{\CTL}[1][]{\lan{CTL}}
\newcommand{\CTLK}[1][]{\lan{CTLK}}

\newcommand{\ATL}[1][]{\lan{ATL_{\stratstyle{#1}}}}

\newcommand{\TCTL}[1][]{\lan{TCTL_{\stratstyle{#1}}}}

\newcommand{\TATL}[1][]{\lan{TATL_{\stratstyle{#1}}}}

\newcommand{\MTL}[1][]{\lan{MTL}}
\newcommand{\MITL}[1][]{\lan{MITL}}
\newcommand{\TPTL}[1][]{\lan{TPTL}}
\newcommand{\SCTL}[1][]{\lan{SCTL_{\stratstyle{#1}}}}
\newcommand{\STCTL}[1][]{\lan{STCTL_{\stratstyle{#1}}}}

\newcommand{\SMTL}[1][]{\lan{SMTL}}
\newcommand{\PTCTL}[1][]{\lan{PTCTL_{\stratstyle{#1}}}}

\newcommand{\PSTCTL}[1][]{\lan{PSTCTL_{\stratstyle{#1}}}}

\newcommand{\PTATL}[1][]{\lan{PTATL_{\stratstyle{#1}}}}

\newcommand{\PATL}[1][]{\lan{PATL_{\stratstyle{#1}}}}

\newcommand{\PCTL}[1][]{\lan{PCTL_{\stratstyle{#1}}}}

\newcommand{\imitator}{IMITATOR\xspace}
\newcommand{\prism}{PRISM\xspace}

\newcommand{\pval}{\ensuremath{z}}
\newcommand{\prob}{\ensuremath{\bowtie}}

\newcommand{\complexityclass}[1]{\ensuremath{\mathbf{{#1}}}\xspace}

\newcommand{\Ptime}{\complexityclass{P}}

\newcommand{\Exptime}{\complexityclass{EXPTIME}}
\newcommand{\EXPTIME}{\Exptime}

\newcommand{\Deltacomplx}[1]{\complexityclass{\Delta_{{#1}}^{\Ptime}}}

\newcommand{\Deltwo}{\Deltacomplx{2}}

\setcopyright{ifaamas}
\acmConference[AAMAS '26]{Proc.\@ of the 25th International Conference
on Autonomous Agents and Multiagent Systems (AAMAS 2026)}{May 25 -- 29, 2026}
{Paphos, Cyprus}{C.~Amato, L.~Dennis, V.~Mascardi, J.~Thangarajah (eds.)}
\copyrightyear{2026}
\acmYear{2026}
\acmDOI{}
\acmPrice{}
\acmISBN{}

\acmSubmissionID{1139}

\title[PSTCTL]{Towards Probabilistic Strategic Timed CTL}

\author{Wojciech Jamroga}
\affiliation{
  \institution{Institute of Computer Science, PAS}
  \city{}
  \country{}}
\affiliation{
  \institution{\& Nicolaus Copernicus University\\in Toruń}
  \city{}
  \country{}}
\email{jamroga@ipipan.waw.pl}

\author{Marta Kwiatkowska}
\affiliation{
  \institution{Institute of Computer Science, PAS}
  \city{}
  \country{}}
\affiliation{
  \institution{\& Department of Computer Science,\\University of Oxford}
  \city{}
  \country{}}
\email{marta.kwiatkowska@cs.ox.ac.uk}

\author{Wojciech Penczek}
\affiliation{
  \institution{Institute of Computer Science,\\Polish Academy of Sciences}
  \city{Warsaw}
  \country{Poland}}
\email{penczek@ipipan.waw.pl}

\author{Laure Petrucci}
\affiliation{
  \institution{LIPN, CNRS UMR 7030,\\Universit\'{e} Sorbonne Paris Nord}
  \city{Villetaneuse}
  \country{France}}
\email{petrucci@lipn.univ-paris13.fr}

\author{Teofil Sidoruk}
\affiliation{
  \institution{Institute of Computer Science,\\Polish Academy of Sciences}
  \city{Warsaw}
  \country{Poland}}
\email{t.sidoruk@ipipan.waw.pl}

\begin{abstract}
We define \PSTCTL, a probabilistic variant of Strategic Timed \CTL (\STCTL), interpreted over stochastic multi-agent systems with continuous time and asynchronous execution semantics. \STCTL extends \TCTL with strategic operators in the style of \ATL. Moreover, we demonstrate the feasibility of verification with \irP-strategies.
\end{abstract}


\newcommand{\BibTeX}{\rm B\kern-.05em{\sc i\kern-.025em b}\kern-.08em\TeX}

\begin{document}

\pagestyle{fancy}
\fancyhead{}

\maketitle 


\section{Introduction}
\label{sec:intro}
Strategic Timed Computation Tree Logic (\STCTL) \cite{AAMAS23STCTL} extends the classical branching-time logic \CTL \cite{Clarke81ctl,Clarke18mcheck2nd} two important directions, firstly by adding discrete- or continuous-time representation, and secondly through 
enabling strategic reasoning in multi-agent systems.
As such, it addresses the ongoing needs and challenges in formal verification, offering a specification language richer than comparable formalisms (\ie, Alternating-time Temporal Logic \ATL \cite{Alur97ATL,Alur02ATL,KacprzakP05,Jamroga15specificationMAS} and its timed variants \cite{Laroussinie06TATL,AAMAS23STCTL}) without creating an overhead in computational complexity.

However, to adequately capture the diversity of real-world behaviours, which are subject to uncertainty, we need to consider probabilistic models and specifications.
Hence, we investigate \PSTCTL, a probabilistic variant of \STCTL.
On one hand, it constitutes 
a strategic extension of the well-known formalisms of \PCTL and \PTCTL \cite{Hansson94logic,BiancoA95,Baier08mcheck}.
In contrast to these two logics, typically considered in synchronous or turn-based settings, \PSTCTL is interpreted over probabilistic continuous-time asynchronous multi-agent systems (\PTAMAS), allowing for reasoning about imperfect information strategies.
On the other hand, \PSTCTL is also a natural branching-time counterpart to the recently proposed linear-time \PTATL \cite{AAMAS25PTATL}, and a missing component in the parallel line of research on branching-time frameworks.
We demonstrate the feasibility of model checking of \PSTCTL 
in practice.
As opposed to recent work on \PTATL \cite{AAMAS25PTATL}, we focus on probabilistic (\irP), rather than deterministic (\irp),  strategies of memoryless agents with imperfect information.

\paragraph*{Related Work.}

\SCTL and \STCTL \cite{AAMAS23STCTL} augment, respectively, the classical branching-time logic \CTL \cite{Clarke81ctl,Clarke18mcheck2nd} and its (continuous or discrete) time extension \TCTL \cite{AlurCD93} with the strategic modality.
\ATL \cite{Alur97ATL,Alur02ATL} and \TATL \cite{Laroussinie06TATL,Knapik19timedATL} are the corresponding linear-time formalisms based on \LTL \cite{Pnueli77temporal}.
\ATL was further developed in multiple directions, \eg by the addition of epistemic reasoning \cite{Guelev12stratcontexts,Jamroga11comparing-ijcai}, strategy contexts \cite{Brihaye09strategycontexts,laroussinie2015augmenting}, or quantification of agents' uncertainty \cite{Tabatabaei23uncertainty}.
The closest linear-time counterpart to \PSTCTL
is the recently proposed \PTATL \cite{AAMAS25PTATL}, notably also in the imperfect information setting (which, due to significantly higher complexity of verification, has seldom been considered in conjunction with probabilistic logics and models).
Untimed \PATL with imperfect information (but interpreted over synchronous models) was theoretically studied in \cite{Belardinelli23probATL,Belardinelli24PATL-probstrats-AAMAS},
and shown to be \Deltwo-complete and in \EXPTIME for deterministic and probabilistic strategies, respectively.

\section{Preliminaries}
\label{sec:preliminaries}
In this section, we recall the relevant theoretical background from \cite{POR4ATL-JAIR,AAMAS23STCTL,AAMAS25PTATL}.
Then, we define the syntax and semantics of \PSTCTL.

\paragraph*{Agents.}
Probabilistic Continuous-time Asynchronous Multi-Agent Systems (\PTAMAS) \cite{AAMAS25PTATL} are networks of Probabilistic Timed Automata (PTA), whose components represent individual agents.
Following the modeling tradition from the theory of concurrent systems, local (private) transitions are asynchronously interleaved, while synchronisations occur on actions shared by two or more agents \cite{Fagin95knowledge,LomuscioPQ10,POR4ATL-JAIR}.
However, the formalism also features an aspect typically present in synchronous frameworks, namely \emph{protocols} that list actions available to agents.
The product of components (or \PTAMAS \emph{model}) captures the system's global behaviour, and the \emph{concrete model} adds continuous-time representation to the states.

\paragraph*{Strategies.}
Conditional plans of an agent (in a coalition), dictating choices in each possible situation, are called (joint) \emph{strategies}
and usually classified \cite{Schobbens04,AAMAS25PTATL} based on the agents' \emph{state information}: perfect~(\stratstyle{I}) vs.~imperfect~(\stratstyle{i}),
\emph{recall of state history}: perfect~(\stratstyle{R}) vs.~no recall~(\stratstyle{r}),
and \emph{action selection}: 
probabilistic~(\stratstyle{P}) vs. deterministic~(\stratstyle{p}).

We focus on memoryless strategies without recall, \ie of type \stratstyle{irP}.
Formally, an \irP-strategy for agent $i$ is a function that maps from $i$'s local states to a probability distribution over $i$'s actions available in these states.
Note that deterministic (\irp) strategies are a special case of \irP where only point distributions are considered.

\paragraph*{Logic.}
The syntax of Probabilistic Strategic Timed \CTL (\PSTCTL):
\begin{center}
$\varphi ::= \prop{p} \mid \neg \varphi \mid \varphi\wedge\varphi \mid \coop{A}\gamma$,
$\gamma ::= \varphi \mid \neg \gamma \mid \gamma\wedge\gamma \mid \Probpath^{\prob\pval}\gamma\Until_I\gamma \mid \Probpath^{\prob\pval}\gamma\Release_I\gamma$,
\end{center}
includes the \emph{strategic modality} $\coop{\A}$
(expressing that agent coalition $\A$ has a strategy to enforce the property that follows),
continuous-time intervals $I\subseteq\mathbb{R}_{0+}$
(restricting the evaluation of temporal operators they are subscribed to),
and the \PCTL-style \emph{probabilistic path operator} $\Probpath^{\prob\pval}$
(stating that the probability of taking a path that satisfies the property that follows
is in relation $\prob$ with constant $\pval$).

Atomic propositions $\prop{p}$, temporal operators $\Until_I,\Release_I$ (and derived $\Sometm_I,\Always_I$), and their Boolean combinations are defined as usual \cite{AAMAS23STCTL,AAMAS25PTATL}.
Below, we give the semantic clause for the case unique to \PSTCTL:

\begin{itemize}
\item $\model, (\State,v) \models \coop{A} \Probpath^{\prob\pval} \psi$ iff there exists a joint strategy $\strat_A$\\
      such that
				    for all $\mu_\model((\State,v),\sigma_\A) \in \outcome_\model((\State,v),\sigma_\A)$
            we have \\ $\mu_\model((\State,v),\sigma_\A)(\{ \seq \mid \model,\seq \models \psi \}) \prob \pval$, where:
						\begin{itemize}
				    \item $\model,\seq \models \gamma_1 U_{I}\gamma_2$ iff 
									there is $r \in I$ such that:   
				          $\model,\pi_\rho(r) \models \gamma_2$ and
				          for all $0\le r' < r$:\  $\model,\pi_\rho(r') \models\gamma_1$.

				    \item $\model,\seq \models \gamma_1 R_{I}\gamma_2$ iff
									for all $r \in I$:   
				          $\model, \pi_\rho(r) \models \gamma_2$ or 
				          there is $0 \le r' < r$:\  $\model,\pi_\rho(r')\models\gamma_1$.
						\end{itemize}
       \end{itemize}
In the above, $\outcome_\model((\State,v),\sigma_\A)$ denotes the \emph{outcome} of strategy $\sigma_\A$ in concrete state $(\State,v)$,
\ie the set of all probability distributions induced by executions from $(\State,v)$ consistent with $\sigma_\A$.
We refer the reader to \cite{AAMAS25PTATL} for formal definitions.
Moreover, $\pi_\seq$ denotes the dense path corresponding to an execution $\seq$, see \cite{AdvancesTimed06}.

\section{Experiments}
\label{sec:expe}
In this section, we discuss practical model checking of \PSTCTL with probabilistic strategies (\PSTCTL[\irP]).
The approach is quite different from the algorithm for \PTATL[\irp], which combined two verifiers: \imitator (handling \irp-strategies and timing constraints) and \prism (handling probabilistic constraints), see \cite{AAMAS25PTATL}.
Since \irP-strategies can choose different actions in each visit of a local state, the strategy encoding introduced in \cite{AAMAS25PTATL} is no longer suitable, and both time and probabilities can be handled by \prism (where remembering the actions chosen for \irp-strategies is not as easy).

To address this issue, for each local state of the coalition agent(s) in the input \PTAMAS $\MASsymb$, we define one parameter per available action, representing the probability of selecting this particular action in the agent's \irP-strategy, see \Cref{fig:ex:prism-params}.

\begin{figure}[H]
\begin{lstlisting}[basicstyle=\scriptsize\ttfamily,keepspaces=true,columns=flexible]
const double p, q, r; // parameters
module Controller
...
  [] locC=0 -> p : (locC'=1) + q : (locC'=2) + r : (locC'=3) + 1-p-q-r : (locC'=4);
\end{lstlisting}
\caption{Probabilistic choice in the \prism model for $n=4$.}
\label{fig:ex:prism-params}
\end{figure}

\paragraph*{Benchmark.}

The classical Train-Gate-Controller (TGC) scenario \cite{AlurFaultyTGC,HoekWooldridge02b,POR4ATL-JAIR} was scaled with the number of trains $n$.
The $i$-th train takes $n+1-i$ time units to go through the tunnel, \eg $n$ for the first one and $1$ for the last.
The formula $\varphi = \coop{C}\Probpath^{\geq 0.8}\Sometm_{[0,T]} (\prop{passed_1} \land \prop{passed_2})$ states that the controller $C$ has a strategy to let the first two trains through the tunnel by time $T$, with a probability of at least $0.8$.
Clearly, it is not satisfied by any \irp-strategy, as opposed to \irP-strategies that assign sufficient probabilities to actions selecting trains 1 and 2 in $C$'s initial state.

Note that, although \prism does not support parametric verification for timed automata models, this issue can be circumvented by extracting the intermediate model generated by the digital clocks engine using the \lstinline[basicstyle=\ttfamily]{-exportdigital} argument.
Following \cite{fmsd06}, under suitable restrictions
this model is a discrete-time MDP that preserves probabilistic reachability values and retains parameter declarations, so can be fed back to \prism.

\paragraph*{Test Platform.}

\prism was executed in the Windows Subsystem for Linux (WSL) on a 4.0 GHz CPU (8 cores, 16 threads) with 64 GB of RAM. Results in \Cref{tab:results} combine verification and model generation time (the latter consistently negligible); \emph{memout} indicates termination of the process upon memory usage exceeding 32 GB (\ie{} half of the total available, following the default WSL setting).

\begin{table}
\centering
\begin{tabular}{ |c|c|c|c| }
 \hline
    n & $T$=5 & $T$=30 & $T$=100 \\
 \hline
	1 train &
    \multicolumn{3}{c|}{n/a (requires $n\geq2$ trains)}  \\
 \hline
	2 trains &
	0.1 s &
	0.1 s &
	0.1 s \\
 \hline
	3 trains &
	0.1 s &
	0.7 s &
	28.6 s \\
 \hline
	4 trains &
	0.4 s &
	4.1 s &
	493 s \\
 \hline
	5 trains &
	4.1 s &
	849 s &
	memout \\
 \hline
\end{tabular}
\vspace{0.2cm}
\caption{Results for TGC with $n$ trains and time constraint $T$.}
\vspace{-0.7cm}
\label{tab:results}
\end{table}

\paragraph*{Results.}

While we were able to successfully verify a strategic property involving the probabilistic choices of an agent for the first time, it is also clear that the scalability of our initial approach is limited.
For more than $n=5$ trains,
\prism no longer returns any synthesised probability values (or their ranges) for the defined parameters.
As such, the cost of encoding probabilistic choices in this manner can be prohibitive for larger models.
However, this is expected, given the theoretical complexity of \PSTCTL verification,
emphasising the need not only for dedicated tools implementing more efficient algorithms, but also complementary techniques such as approximations and model abstractions.

\section{Conclusions}
\label{sec:conclu}
We proposed \PSTCTL, a probabilistic variant of strategic \CTL with continuous time,
and tackled for the first time the case of \irP-strategies in practical model checking.
There are several directions for further research,
including formally establishing the expressive power and model checking complexity of \PSTCTL[\irP] and \PSTCTL[\irp],
as well as investigating timed vs. untimed strategies.


\begin{acks}
This work was supported
by CNRS IRP ``Le Tr\'{o}jk{\k a}t'',
by NCBR Poland \& FNR Luxembourg under the PolLux/FNR-CORE project SpaceVote (POLLUX-XI/14/SpaceVote/2023),
by the PHC Polonium project MoCcA (BPN/BFR/2023/1/00045),
and by the ANR-22-CE48-0012 project BISOUS.
M. Kwiatkowska contributed while on sabbatical and acknowledges funding from the ERC under the European Union’s Horizon 2020 research and innovation programme (FUN2MODEL, grant agreement No.~834115).
\end{acks}


\balance
\bibliographystyle{ACM-Reference-Format} 
\bibliography{report,wojtek,wojtek-own}


\end{document}